\documentstyle[preprint,pre,aps]{revtex}  
\input epsf.sty 
\begin{document}  
\draft  
\title{Magnetic history dependence of metastable states 
in systems with dipolar interactions.
}  
\author{Oscar Iglesias and Am\'{\i}lcar Labarta}   
\address{  
Department de F\'{\i}sica Fonamental, Facultat de F\'{\i}sica, Universitat  
de Barcelona, Diagonal 647, 08028 Barcelona, Spain}  
\date{\today}  
\maketitle  
\newcommand{\svar}{T \ln(t/\tau_0)} 
\newcommand{\phrb}{Phys. Rev. B } 
\newcommand{\phrl}{ Phys. Rev. Lett. } 
\newcommand{\phre}{Phys. Rev. E } 
\newcommand{\jm}{J. Magn. Magn. Mater. } 
\newcommand{\psa}{Phys. Stat. Sol. A }  
\newcommand{\jap}{J. Appl. Phys. }  
\newcommand{\beq}{\begin{equation}} 
\newcommand{\eeq}{\end{equation}} 
\newcommand{\bea}{\begin{eqnarray}} 
\newcommand{\eea}{\end{eqnarray}} 
\begin{abstract}  
We present the results of a Monte Carlo simulation of the ground state and
magnetic relaxation of a model of a thin film consisting on a
two-dimensional square lattice of Heisenberg spins with perpendicular
anisotropy K, exchange J and long-range dipolar interactions $g$. We have
studied the ground state configurations of this system for a wide range of
the interaction parameters J/g, K/g by means of the simulated annealing
procedure, showing that the model is able to reproduce the different
magnetic configurations found in real samples. We have found the existence
of a certain range of K/g, J/g values for which in-plane and out-of-plane
configurations are quasi-degenerated in energy. We show that when a system in
this region of parameters is perturbed by an external force that is
subsequently removed different kinds of ordering may be induced depending
on the followed procedure. In particular, simulations of relaxations
from saturation under an a.c. demagnetizing field or in zero field are in
qualitative agreement with recent experiments on epitaxial and granular
alloy thin films, which show a wide variety of magnetic patterns depending
on their magnetic history.
\end{abstract}  
\section{Introduction}
\label{SecI}
The magnetic and transport properties of thin films and nanostructured materials
have been the subject of intense research during at least the last two past 
decades because of their interest for magnetic recording and technological 
applications \cite{Magnetoelectronics}. Among the interesting phenomena observed in this kind of materials
there are the reorientation transitions of the magnetization with increasing 
temperature \cite{Allenspachprl92,Webbprl93,Macprl98} or film thickness 
\cite{Allenspachprl92,Allenspachprl90,Hehnprb96}, 
the giant magnetoresistance effect \cite{Berkowitprl92,Xiaoprl92,Kentjap99}, 
and the wide variety of magnetic 
patterns that can be stabilized depending on the interplay between the 
perpendicular induced surface anisotropy, the exchange interaction and 
the long-range dipolar forces between the microscopic entities. 

The development of different microscopy techniques like MFM, AFM, LEM or SEMPA 
\cite{Danjm99} have helped experimentalists to understand the interplay between 
the transport properties and the magnetic structure of the materials. Moreover, 
these techniques have shown that the magnetic structures formed in thin films
are strongly influenced by the magnetic history of the sample. Thus, the
observed patterns may display either out-of-plane ordering with labyrinthian 
striped domains \cite{Francoie98,Francoejpb99} or bubble-like patterns 
\cite{Hehnprb96,Greggprl96} as well as in-plane vortex-like structures 
\cite{Donnetjpd95}.  

Most of the experimental samples being studied at present are epitaxial
magnetic thin films of thickness ranging from $10$ to $500$ nm 
\cite{Hehnprb96,Donnetjpd95,Stampsjap97,Bochiprl95,Hehnjm97}
or granular alloy films from $200$ nm to $1 \mu$m thickness 
\cite{Garvinapl95,Garvinjap96}.
Current experimental techniques allow to control the ferromagnetic content 
of granular alloys deposited on nonmagnetic metallic matrices \cite{Francoejpb99} 
and therefore vary the range of dipolar and exchange interactions between the 
grains. It seems clear that these experimental systems should be efficiently 
modeled by a two-dimensional lattice of spins coupled by exchange interaction and  
shape anisotropy perpendicular to the plane of the sample and that the  
consideration of long-range forces between the spins, mainly of dipolar origin,
is essential to understand the magnetic properties of these systems. 
The finite thickness of the samples can be taken into account by decreasing 
the value of the effective local anisotropy constant.   
 
From the theoretical point of view the current understanding  
of dipolar spin systems can be summarized as follows.  
It is well established that the ground state of a pure planar  
dipolar system (no anisotropy and no exchange interaction)  
is a continuously degenerate manifold of antiferromagnetic  
(AF) states \cite{Macprb97,Prakashprb90} (checkerboard phase). The introduction of exchange 
interactions ($J>0$) between the spins establishes a competition between 
the short-range ferromagnetic (FM) order and the long-range dipolar 
interaction that favours AF order \cite{Macprb95,Macprl98}.  
Contrary to what one would expect, increasing $J$ does not serve to 
stabilize a FM ground state but instead results in the appearance of 
striped phases of increasing width that do not disappear at high $J$. 
A finite perpendicular anisotropy $K$ favours  
the formation of out-of-plane configurations against the in-plane  
configurations induced by the dipolar interaction.     
Monte Carlo simulations \cite{Macprl98,Macprl96,Chuiprb94,Chuiprb95,Chuiprl95} 
as well as theoretical analysis 
\cite{Prakashprb90,Yafetjap86,Yafetprb88,Kaplanjm93,Taylorjpc93} 
have shown that there is a reorientation transition 
from out to in-plane order or from in to out-of-plane order depending on the 
ratio of $K$ to $J$ as the temperature is increased \cite{Macprl98,Macprl96}. 
In the above mentioned works main attention was put on the determination of 
phase diagrams but apparently an accurate description of the detailed structure of the 
ground state has not been reported. 

In this work we present the results of extensive Monte Carlo simulations of a 
model of a thin film with the aim to explain the variety of magnetic behaviours 
of the above mentioned experimental observations. We start by the description 
of the model Hamiltonian Sec. \ref{SecII}, in Sec. \ref{SecIII} the phase diagram
and ground state configurations of the model are presented demonstrating that
it qualitatively reproduces the patterns observed in experiments. In Sec. 
\ref{SecIV} we present the results of two simulations that show the effect of 
two different magnetic histories on the magnetic order of the system. 
We conclude with the conclusions in Sec. \ref{SecV}

\section{Numerical Model and Hamiltonian}
\label{SecII}

Our general model of a thin film consists of a two-dimensional 
square lattice (lattice spacing $a$ and linear size $N$) 
of continuous spins Heisenberg ${\bf S}_i$  
with magnetic moment $\mu$ and uniaxial anisotropy $K$ perpendicular  
to the lattice plane described by the Hamiltonian  
\beq 
\label{Ham1} 
{\cal H}=-{\bar J} {\cal H}_{exch}-{\bar K}{\cal H}_{anis}+ {\cal H}_{dip} 
\eeq 
where ${\bar J}$, ${\bar K}$, and ${\cal H}$ are given in units of  
\beq 
g\equiv \frac{\mu^2}{a^3} \nonumber 
\eeq 
with $a$ the lattice spacing. The spin ${\bf S}_i$ may represent either an 
atomic spin or the total spin of a grain.
 
The three terms in the Hamiltonian correspond to short-range nearest 
neighbour exchange interaction (direct between atomic spins or indirect through the
matrix), uniaxial anisotropy energy perpendicular to the lattice plane, 
and long-range dipolar interaction 
\bea 
\label{Ham2} 
&&{\cal H}_{exch} = \sum_{n.n.} ({\bf S}_i \cdot {\bf S}_j) \\ \nonumber 
&&{\cal H}_{anis} = \sum_{n=1}^{N^2} \, (S_n^z)^2 \  \\ \nonumber 
&&{\cal H}_{dip} = \sum_{n\neq m}^{N^2} \, \sum_{\alpha,\beta=1}^{3} \,S_n^\alpha  
                 \, W_{nm}^{(\alpha \beta)}  \,S_m^\beta \ . 
\eea 

In the last term we have defined the following set of dipolar interaction matrices  
\beq 
\label{Dipmat} 
W_{nm}^{(\alpha \beta)}= 
\frac {1}{r_{nm}^3} 
\left( 
\delta_{\alpha\beta} - {3 \delta_{\alpha\gamma} \delta_{\beta\eta}  
r_{nm}^\gamma r_{nm}^\eta \over  
r_{nm}^2} 
\right) \ ,  
\eeq 
$r_{nm}$ a vector connecting spins at sites $n$ and $m$. 
The matrices $W_{nm}^{(\alpha \beta)}$ depend only on the lattice geometry and boundary 
conditions and not on the particular spin configuration. In this expression the first 
term favours antiferromagnetic long-range order while the second one introduces an 
effective easy-plane anisotropy. 
 
Thus, the properties of the model depend only on two parameters. $\bar J$, 
which accounts for the competition between the ferromagnetic ($J>0$)
interaction and the antiferromagnetic order induced by $g$, 
and $\bar K$, which accounts for the competition between the out-of-plane 
alignment favoured by $K$ and the in-plane order induced by $g$. 
 
This model reduces to the {\sl Ising model} in the limit $K=+\infty$  
when the out-of-plane components are restricted to $S_i^{z}=\pm 1$,   
and to the {\sl planar model} when $K=-\infty$ and the spins are forced  
to lie in the lattice plane. Both cases depend only on one parameter 
$J/g$, the ratio of exchange to dipolar energies and are described by 
the same formal Hamiltonian ${\cal H}=-{\bar J}{\cal H}_{exch}+ 
{\cal H}_{dip}$. All the simulations have been performed on a system
of size $50\times 50$ with periodic boundary conditions. 

\section{Ground State properties}
\label{SecIII}
We start by studying the properties of the ground state of the model.
For this purpose we have followed the simulated thermal annealing method:
starting at a high temperature with a configuration with spins randomly 
oriented, the temperature is slowly decreased by a constant factor $\Delta T$
(we started with a temperature of 10 and we used a reduction factor of $0.9$),
at every temperature step the system is allowed to evolve during a number $t$ of 
MC steps long enough (usually between 200 and 250 MCS per spin in our system) 
so as to reach thermal equilibrium at every temperature, the process is continued
until the number of accepted trial jumps is less than a small percentage.  
In this way we have obtained the ground state energies and configurations of
the system.

\subsection{Phase diagram}

In Fig. \ref{WS1_fig} we present  
the results of the simulations for the ground state energies of the finite 
$K$ model as a function of the reduced exchange parameter $\bar J$ and for 
different values of the anisotropy constant (open symbols). 
In filled circles the energy of the corresponding Ising model ($K=+ \infty$) 
is also given for comparison, the dashed lines correspond to the same curve 
with the corrections for the different finite anisotropy values added. 
The continuous line corresponds to the same calculation for the planar (XY) 
model. 

A characteristic feature of the finite $\bar K$ model is that it
behaves in a bimodal way. For small $\bar J$ ($\bar J<\bar J^{\star}$)
it is completely equivalent to the corresponding Ising model, displaying
out-of-plane order while for $\bar J>\bar J^{\star}$ it behaves like 
the planar model and it orders in-plane. The value of $\bar J^{\star}$ at which the 
crossover occurs is simply the one for which the ground state energy of the 
planar model (thick solid line) equals the energy of the corresponding Ising 
model with the finite anisotropy correction corresponding to a given 
value of $K$ (dashed lines). Therefore, the phase diagram for the finite $K$ 
model can be directly obtained from the results for the planar model by 
shifting the energy of the Ising model with the corresponding value of $K$.
Consequently, the Heisenberg model for finite $K$ is the combination
of the simpler Ising and planar models and its behaviour is dominated
by the one having the lowest ground state energy. 

The region of parameters of interest to us is precisely those values of
$\bar J$ around the crossover points $\bar J^{\star}$ since in this region the 
out and in-plane configurations are quasi-degenerated in energy and the system
displays metastability, that is to say, different kinds of ordering may be
induced depending on how the system is driven to the quasi-equilibrium 
state as we will show in the next sections.

\subsection{Configurations}

Before proceeding further let us analyze the ground state configurations
in more detail. On the one hand, for $\bar J<\bar J^{\star}$ (Ising regime) 
all the curves show three characteristic regions corresponding 
to different kinds of-out-of plane order. For small values of $\bar J$ the 
dipolar energy dominates over the exchange energy and the system orders 
antiferromagnetically (AF) in a checkerboarded phase of increasing energy as 
$\bar J$ increases (two left columns in Fig. \ref{WS2_fig}). 
At intermediate $\bar J$ values the system enters a 
constant energy region in which a phase of AF stripes of width $h=1$ is 
stabilized by the exchange interaction (third column from the left in Fig. 
\ref{WS2_fig}). Out from this plateau the FM ordering increases resulting 
in a widening of the stripes and an almost linear decrease of the energy 
with $\bar J$ (forth and fifth columns for $\bar K= 5.0, 10.0$ in the same 
figure).

In the metastable region ($\bar J\simeq \bar J^{\star}$) the system starts to
turn to the planar regime first displaying configurations with a regular 
array of vortices with remanent out of plane internal order. This fact 
can be observed for the case $\bar K= 3.3, \bar J= 1, 1.33$ in Fig.
\ref{WS2_fig} and Fig. \ref{WS3_fig} where the in-plane projections of the
spins are displayed with arrows. The diameter of the vortices progressively 
increases with increasing $\bar J$ until the FM order induced by the 
exchange energy stabilizes the in-plane FM configuration.

Moreover, at small values of $\bar K$, the system can change to the planar 
phase without entering some or any of the above mentioned Ising regions.
On the contrary, at high enough $\bar K$ the system never crosses to
the planar phase, behaving as the Ising model for any value of $\bar J$. 
\section{Magnetic history dependence simulations}
\label{SecIV}

In this section we will show simulations that mimic some of
the experimental processes found in recent experiments on thin films. 
In all of them the measuring protocol consists in taking the sample out
of its equilibrium state by the application of an external perturbation 
that is subsequently removed. Usually this external perturbation is
a magnetic field applied perpendicular to or in the thin film plane.
The final state induced by this procedure may be very different from
the initial one, magnetic domains may be erased or created 
depending on the magnetic history to which the sample has been submitted.

\subsection{Relaxation after perpendicular saturation}

In the first simulation experiment we have considered a system with 
$\bar K= 3.3$ and $\bar J= 1.0$ which is just at the crossover between
the Ising and planar regimes and has a ground state configuration with
in-plane vortices. We simulate the application of a saturating field 
perpendicular to the film plane by starting with a configuration with
all the spins pointing along the positive easy-axis direction $S_i^z= + 1$
and we let the system relax to equilibrium in zero applied field during
a long sequence of Monte Carlo steps during which we record the intermediate
system configurations. One example of temporal evolution is shown in 
Fig. \ref{WS4_fig} where we can see a sequence of snapshots of the system taken 
at different stages of the relaxation. In the first stages of the evolution
the system forms striped out-of-plane structures that evolve very slowly
towards the equilibrium configuration of in-plane vortices. Similar results 
are obtained for other values of $\bar J$ close to the intersection point 
of the Ising and planar regimes. For values of $\bar J$ smaller than 
$\bar J^{\star}$ the system reaches the same state as after an annealing process.

Defects and imperfections in real thin films (commonly present in granular 
alloys) may act as pinning centers of these intermediate structures.
This may be the explanation of the change observed in MFM
images of granular films with low concentration of the FM content 
\cite{Hehnprb96,Francoejpb99,Greggprl96,Hehnjm97} that in
the virgin state show no out-of-plane order but after the application
of $10 kOe$ perpendicular to the film plane display striped and bubble-like
domains similar to the ones obtained in the simulation.

\subsection{Demagnetizing filed cycling}

The second kind of process consists in the application of a demagnetizing field 
cycling in the perpendicular direction. 
The parameters used for the simulation are
in this case $\bar K= 5.0, \bar J= 2.1$, also in the metastable region.
As in the previous case in the initial state the system is saturated in
positive perpendicular direction with a field $H_0$ but now the field is 
cycled from the positive to negative direction and progressively reduced
in magnitude with a period $T$ (see the drawing at the top of Fig. 
\ref{WS5_fig}). In the simulation the initial field has been chosen as 
the minimum allowing the system to escape from the initial
saturated state ($H_0= 10.4$) and the period ($T= 40 MCS$) is such that 
the system has time to reverse its magnetization during the reversal of the field.

The results are displayed in the sequence of snapshots of Fig. \ref{WS5_fig}.
At the first stages of the process (not shown in the figure) the spins reverse
following at each reversal of the field. As the time elapses and the field 
decreases some reversed groups of spins start to nucleate (black spots in
the two first rows). They continue to grow forming out-of-plane labyrinthian
configurations separated by in-plane ordered zones (grey areas) arranged in 
vortices. As in the preceding experiment we find that a system with an in-plane 
ordered ground state may be driven to a very different ordering state by the
magnetic history. What is more remarkable in this case is that the state attained 
after the cycling process it is not lost with time. Far from relaxing to the in-plane
ordered ground state, when the system is allowed to relax in zero field, 
the incipient structure formed during the demagnetizing cycle is stabilized.
The narrow adjacent stripes coalesce one with another to form wider stripes 
separated by narrower regions of in-plane spins.

\section{Conclusions}
\label{SecV}

We have shown that a model of two dimensional Heisenberg spins with anisotropy
perependicular to the plane and interacting via exchange and long-range dipolar 
forces is able to reproduce the different magnetic patterns observed in experiments, 
from out-of-plane labyrinthian and striped domains to in-plane FM and vortex 
structures. The long-range character of the dipolar interaction play an essential 
role to understand the ground state properties of this system, the results 
would have been very different if the dipolar field acting on the spins had
been replaced by a mean-field demagnetizing field as is usually done in some
works. An interesting characteristic of the model is that it behaves
as the limiting Ising of planar models depending on the values of $\bar J$ for
a given value of $\bar K$. However, for values of $\bar J$ around the itersection
between the Ising and the planar ground state lines, in-plane and out-of-plane
configurations are quasidegenerated in energy and metastable. In this range
of parameters our simulations are able to reproduce a surprising experimental 
observation \cite{Hehnprb96,Francoie98,Francoejpb99,Greggprl96,Hehnjm97}: 
if a magnetic field perpendicular 
to the film plane is applied to a virgin sample with in-plane FM domains, the 
out-of plane component of the magnetization increases by a factor of 10 and the
magnetic pattern displays well contrasted domains. A similar situation happens
after perpedicular demagnetizing cycles, now the domains elongate and become wider.
The last case could be thought as dinamical phase transition from in to out-of-plane
order induced by a driving time dependent magnetic field similar to that
observed for Ising spins \cite{Sidesprl98,Acharyyaepjb98,Misrapre98}. Therefore, 
the application of an external perturbation that changes momentarily the energy 
landscape together with the existence of highly metastable states facilitates the
driving of the system to a new stable configuration that nonetheless it is
not the equilibrium one.
\section*{Acknowledgements}
We acknowledge CESCA and CEPBA under coordination of $C^{4}$ for the computer 
facilities. This work was supported by CICYT through project MAT97-0404 and
CIRIT under project SR-119. 



\begin{figure} 
\caption{Ground state energy as a function of the reduced exchange constant 
$\bar J=J/g$ for different values of the anisotropy constant $\bar K=K/g$
as indicated in the figure. Open symbols correspond to the general finite
anisotropy case described by the Hamiltonian (\ref{Ham1}). Filled circles
correspond to the Ising model ($K=+\infty$) and the continuous line to the 
planar (XY) model ($K=-\infty$). The dashed lines show 
the ground state energy of the Ising model plus the anisotropy energy 
corresponding to the nearest finite $\bar K$ curve. 
}
\label{WS1_fig} 
\end{figure} 

\begin{figure} 
\caption{Ground state configurations obtained by the method of simulated 
annealing for the finite $K$ model for a system of $50\times 50$ spins and 
some of the $\bar J$ and $\bar K$ values of the phase diagram of Fig.
\ref{WS1_fig}. Only the out-of-plane projections ($S_z$) are shown
in a grey scale ranging from white ($S_z=+1$) to black ($S_z=-1$).  
} 
\label{WS2_fig} 
\end{figure} 
\begin{figure} 
\caption{In-plane projections of some of the configurations shown in the
previous figure \ref{WS2_fig} are indicated by arrows. 
} 
\label{WS3_fig} 
\end{figure} 
\begin{figure} 
\caption{Snapshots of the configurations at different stages of the 
relaxation in zero field of an initially saturated sample with 
$K/g= 3.3, J/g=1.0$ and spins pointing up along the anisotropy easy axis. 
Images in the right column indicate the out-of-plane component of the 
magnetization (ranging from white for up spins to black fot down spins)
while the left column corresponds to the in-plane projections of a 
detail of a $25\times 25$ region in the upper left corner of a 
$50\times 50$ system.
} 
\label{WS4_fig} 
\end{figure} 
\begin{figure} 
\caption{Snapshots of the configurations at different stages of 
a demagnetizing field cycling in the perpendicular direction applied to
an initially saturated sample with $K/g= 5.0, J/g= 2.1$. 
The drawing on the top shows the variation of the field with time, for
this simulation the initial magnitude of the field is $H_0= 10.4$ and
its period $T= 40 MCS$. Left and right columns correspond to the in and 
out-of-plane projections of the spins in the same way that the preceding figure.
The first five rows show snapshots during the field cycling. The configuration 
at the bottom is the one attained after relaxation in zero field. 
} 
\label{WS5_fig} 
\end{figure} 

\end{document}